\documentclass[a4paper,11pt,final]{article}

\usepackage{authblk}

\usepackage{graphics}
\usepackage{graphicx}
\usepackage{epsfig}

\usepackage{amsthm}

\usepackage{amsmath}
\usepackage{amssymb}

\date{}

\begin{document}

\title{Multiphase Partitions of Lattice  Random Walks}

\author[1]{Massimiliano Giona$^*$}
\author[1]{Davide Cocco}
\affil[1]{Dipartimento di Ingegneria Chimica DICMA
Facolt\`{a} di Ingegneria, La Sapienza Universit\`{a} di Roma
via Eudossiana 18, 00184, Roma, Italy  \authorcr
$^*$  Email: massimiliano.giona@uniroma1.it}

\maketitle

\begin{abstract}
Considering the dynamics of non-interacting particles randomly
moving on a lattice, the occurrence of a discontinuous transition
in the values of the lattice parameters (lattice spacing and hopping times)
determines  the uprisal of two lattice phases.
In this Letter we show that the hyperbolic hydrodynamic model
obtained by enforcing the boundedness of lattice  velocities
derived in \cite{giona_lrw} correctly describes the dynamics
of the system and permits to derive easily the
boundary condition at the interface, which, contrarily to the
common belief, involves the lattice velocities in the two phases
and not the phase diffusivities.
The dispersion properties  of independent particles moving
on an infinite lattice composed by the periodic
repetition of a multiphase unit cell are investigated.
It is shown that the hyperbolic transport theory correctly
predicts the effective diffusion coefficient over all
the range of parameter values, while the corresponding continuous
parabolic models deriving from Langevin equations for particle
motion fail. The failure of parabolic transport models
is shown via a simple  numerical experiment. 
\end{abstract}

Lattice models of particle dynamics represent a robust  conceptual
backbone in
statistical theory of non-equilibrium processes,
 finding  broad and diversified applications in practically all the
branches of physics \cite{gen1,gen2}. They constitute
a simple {\em gedanke} experimental enviroment in order to
derive, from simple local interaction rules, the corresponding
hydrodynamic models in a continuous  space-time setting 
\cite{contlim1,contlim2}.

Even in the case of systems of noninteracting particles,
a rich variety of possible phenomenologies
arises, associated with lattice heterogeneities and
impurities \cite{lax},
randomly distributed multisite structures \cite{multisite,weiss}, disorder, 
percolation
and phase transitions \cite{disorder}, anomalous behavior induced
by a continuous distribution of hopping times and
hopping lengths (that can be treated within the framework of
Continuous Time Random Walk) \cite{ctrw1,ctrw2}, etc.

In recent years, lattice heterogeneity has been studied
in connection with infiltration dynamics, and solute
partition in two lattice phases, defined by the
decomposition of the lattice in two subsets possessing
different lattice parameters \cite{kora1,kora2,dbc1,dbc2,dbc3}.
The latter problem has great current interest in biological
applications involving active swimmers moving in nonuniform
fields modulating their mobilities \cite{swimmers},
and is connected to fundamental problems involving 
 the stochastic modeling of non-equilibrium phenomena,
associated with the proper choice of the most suitable
stochastic calculus (Ito, Stratonovich, H\"anggi-Klimontovich)
\cite{ito1,ito2}, and with the properties
of the  equilibrium invariant densities and their connection with
local transport properties \cite{equilibrium}.

Although the current research focus is mostly oriented towards particle
motion determining the occurrence  of anomalous diffusive phenomena
\cite{kora2}, the case of the simplest possible model
of lattice random walk involving noninteracting particles has
shown that 
some interesting properties are still to be unveiled, especially
as  regards its continuous hydrodynamic description.
It has been shown recently in \cite{giona_lrw} that
the the classical lattice random walk of independent
particles can be described in a continuous space-time setting
by means of a hyperbolic transport model analogous to
those arising in the theory of Generalized Poisson-Kac processes
\cite{gpk0,gpk1}.
The hyperbolic transport model accounts intrinsically for the 
finite propagation
velocity of the process, and provides an accurate quantitative
description not only of the long-term properties but also
of the initial stages of the dynamics \cite{giona_lrw}.
 In the case of a symmetric lattice random walk,
 defined by the characteristic
site distance $\delta$ and  hopping time $\tau$
between nearest neighboring sites, the  hyperbolic continuous model
involves the partial probabilities $\{ p_\pm(x,t) \}$,
where $p_{\pm}(x,t)\, dx$ is the fraction of particles
in the spatial interval $(x,x+dx)$ at time $t$ moving towards the
right ($p_+$) or to the left ($p_-$), satisfying the 
equations
\begin{equation}
\frac{\partial p_\pm(x,t)}{\partial t}= \mp b \, \frac{\partial p_\pm(x,t)}{\partial x}
\mp \lambda \,   \left [p_+(x,t) - p_-(x,t) \right ]
\label{eq1}
\end{equation}
where $b=\delta/\tau$, and $\lambda=1/\tau$ (see the Appendix).
The overall probability density  of the process is $p(x,t)=p_+(x,t)+p_-(x,t)$,
and the associated flux  $J(x,t)=b \, [p_+(x,t)-p_-(x,t)]$.

\begin{figure}
\begin{center}
\includegraphics[width=8cm]{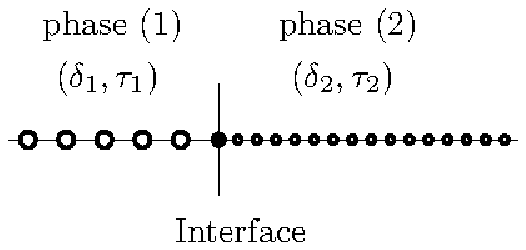}
\end{center}
\caption{Schematic representation of a MuPh-LRW.}
\label{Fig1}
\end{figure}

In this work, we consider  the multiphase extension of the symmetric
random walk, as depicted in figure \ref{Fig1}. The MultiPhase Lattice
Random Walk, henceforth MuPh-LRW for short, is a simple
random walk on a lattice ${\mathbb Z}$, in which the physical
lattice parameters $(\delta,\tau)$ admit a sudden transition
at some lattice point, say $z_0 \in {\mathbb Z}$ so that
$(\delta_1,\tau_1)$ holds for $z<z_0$, and $(\delta_2,\tau_2)$
for $z > z_0$. The lattice point $z_0$ represents the interfase
separating the two lattice phases.
Within  each lattice phase particle motion is a  symmetric LRW,
corresponding, in the continuous limit, to an emergent purely diffusive 
behavior defined by the phase diffusivities $D_h=\delta_h^2/2 \tau_h$, 
$h=1,2$.
It remains to specify the motion at the
interfacial point $z_0$. Two cases can occur: (i) if the
interfacial point is ``neutral'' with respect to phase selection,
so that equal probabilities characterize the jump from $z_0$
to one of its two nearest neighboring sites,
 the interface is referred to
as {\em ideal}; (ii) if the  probabilities of moving
from the interface towards the sites of one of the two phases
are different, the interface exerts a specific and active
selection, and it will be referred to as {\em non ideal} or {\em active}.
In this Letter we focus exclusively on ideal interfaces.

In a discrete space-time description, the MuPh-LRW corresponds to
a simple symmetric LRW defined by the dynamics $z_{n+1}=z_n\pm 1
\; \mbox{Prob}. \; 1/2$, where $n=0,1,\dots$ is the lattice
time. In a physical setting, indicating with $x_n$ the particle
spatial position and $t_n$ the physical time, MuPh-LRW corresponds
to a subordination of the stochastic lattice motion according to
phase heterogeneity. More precisely, let $\sigma_n=\sigma(x_n)$
be the phase-function, $\sigma_n=\{1,2,0\}$, depending whether
the site $x_n$ belongs to phase ``1'', ``2'', or is an interfacial
site (``0'')., the MuPh-LRW dynamics in the presence of
ideal interfaces is defined by
\begin{eqnarray}
x_{n+1} & = & x_n  + \mbox{sgn}(x_{n+1}-x_n) \, M[\sigma_n,\sigma_{n \pm 1}] \;\;\;
\mbox{Prob}. \; 1/2 \nonumber \\
t_{n+1} & = & t_n+T[\sigma_n,\sigma_{n+1}]
\label{eq2}
\end{eqnarray}
where $\mbox{sgn}(\cdot)$ is the sign function, and 
in which the symmetric matrices $M[s_1,s_2]$, $T[s_1,s_2]$
are defined by the transport (metric and temporal) properties
of the two phases, $M[h,h]=M[0,h]=M[h,0]=\delta_h$,
$T[h,h]=T[0,h]=T[h,0]=\tau_h$, $h=1,2$.

Two main questions arise: (i) the definition of a continuous hydrodynamic
model for MuPh-LRW, and (ii), strictly connected to (i), the
assessment of the proper boundary condition at an ideal interface
in a continuous setting of the dynamics. These two issues
are closely related to each other. As regards the hydrodynamic
description, the hyperbolic approach introduced in \cite{giona_lrw}
can be applied to each phase. This corresponds to
consider  eq. (\ref{eq1})  for each phase, with $p_\pm(x,t)$ substituted
by the phase partial concentration $p_\pm^{(h)}(x,t)$, defined
for $x$ within each disjoint domain of definition of the phases,
and $b$ and $\lambda$ with $b_h=\delta_h/\tau_h$,
$\lambda_h=1/\tau_h$. This follows also from eq. (\ref{eq2})
by considering the subordination of the physical time $t$
with respect to the lattice time $n$ for processes
possessing finite propagation velocity (Poisson-Kac processes)
(see the Appendix). In the case of ideal interfaces, there
is no active effect of the interface on the partition
of solute particles in the two phases, and the hyperbolic
approach    based on eq. (\ref{eq1})
implies the continuity of the partial fluxes across
an ideal interface located at $x_0$ (corresponding to the
lattice coordinate $z_0$),
\begin{equation}
\left . b_1  \, p_\pm^{(1)}(x,t) \right |_{x=x_0}
= \left . b_2  \, p_\pm^{(2)}(x,t) \right |_{x=x_0}
\label{eq3}
\end{equation}
This condition can be further justified by enforcing the
lattice representation of the dynamics, consistently with
the analysis developed in the Appendix.
Eq. (\ref{eq3}) obviously predicts the continuity of
the normal  fluxes at the interface, $J^{(1)}(x_0,t)=J^{(2)}(x_0,t)$,
and the boundary condition for the overall particle density 
\begin{equation}
\left . \frac{p^{(2)}(x,t)}{p^{(1)}(x,t)} \right |_{x=x_0}= \frac{b_1}{b_2}
= \frac{\delta_1 \, \tau_2}{\delta_2 \, \tau_1}
\label{eq4}
\end{equation}
In point of fact, eq. (\ref{eq4}) represents a change of paradigm with
respect to the usual approach to boundary conditions applied at interfaces
in the presence of diffusion, in which $p^{(2)}/p^{(1)}|_{x=x_0}$
is assumed equal to the ratio of the diffusivities $D_1/D_2$
\cite{kora1,dbc1,dbc2,dbc3}. Eq. (\ref{eq3}) finds its natural
explanation  in the stochastic models in which the finite
value of the propagation velocities $b_h$ is assumed, while
it is ``alien'' to the classical parabolic approach.
In this framework, the analysis of MuPh-LRW is a significant
benchmark to test the importance of the physical assumptions
underlying hyperbolic transport theories \cite{gpk1,gpk2,gpk3}.

Direct numerical simulations of MuPh-LRW provides a clear answer to this
question. Consider a MuPh-LRW on a closed domain
$x \in [-1,1]$ equipped with zero-flux conditions at the
endpoints. The interval $[-1,0)$ corresponds to the
lattice phase ``1'', $(0,1]$ to the lattice
phase ``2'', and the
interface is located at $x=x_0=0$. In the simulations,
$\delta_1=1/N$, $N_1=N$ is the number of lattice sites in phase ``1'',
$\delta_2=\delta_1/\alpha$, where $\alpha$ is an integer, so that
$N_2=\alpha \, N$ is the number of sites of phase ``2'',
 while $\tau_1=1$ and $\tau_2$ freely varies.
An ensemble of $10^6$ particles  is considered, initially located
at the interface.

Figures 2 and  \ref{Fig3} depict the comparison of the
lattice simulations of MuPh-LRW with the results obtained by
integrating the hyperbolic equations (\ref{eq1}) for each phase
and in each disjoint phase domain where the boundary conditions
\ref{eq3} have been applied at the interface $x=0$.
Figure 2 refers to $\delta_2=\delta_1/4$,
$\tau_2=\tau_1$, so that $p^{(2)}/p^{(1)}|_{x=0}=4$, while
the classical diffusive boundary condition provides
$p^{(2)}/p^{(1)}=16$. Figure \ref{Fig3} refers to $\delta_2=\delta_1/2$,
$\tau_2=\tau_1/2$, at which the hyperbolic theory predicts
a smooth overall concentration profile across the
interface as $p^{(2)}/p^{(1)}|_{x=0}=1$, while the diffusive boundary condition
implies a discontinuity  $p^{(2)}/p^{(1)}|_{x=0}=2$.

\begin{figure}
\begin{center}
\includegraphics[width=14cm]{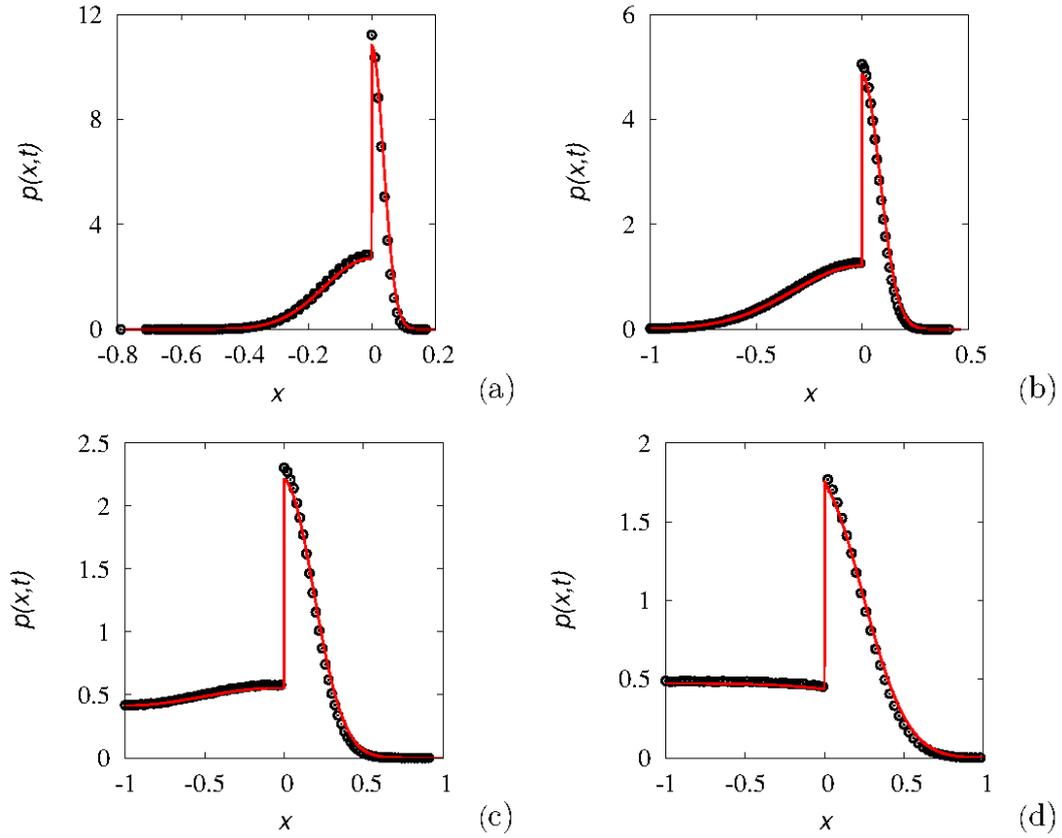}
\caption{Probability density function $p(x,t)$ vs $x$ at several
time instants for
 $\delta_2=\delta_1/4$, $\tau_2=\tau_1$.
Symbols ($\circ$) are the results of lattice simulations, continuous
lines the solutions of the corresponding hyperbolic model.
Panel (a) refers to $t=2 \times 10^2$, (b) to $t=10^3$, (c) to
$t=5 \times 10^3$, (d) to $t=10^4$.}
\end{center}
\label{xFig2}
\end{figure}

\begin{figure}
\begin{center}
\includegraphics[width=14cm]{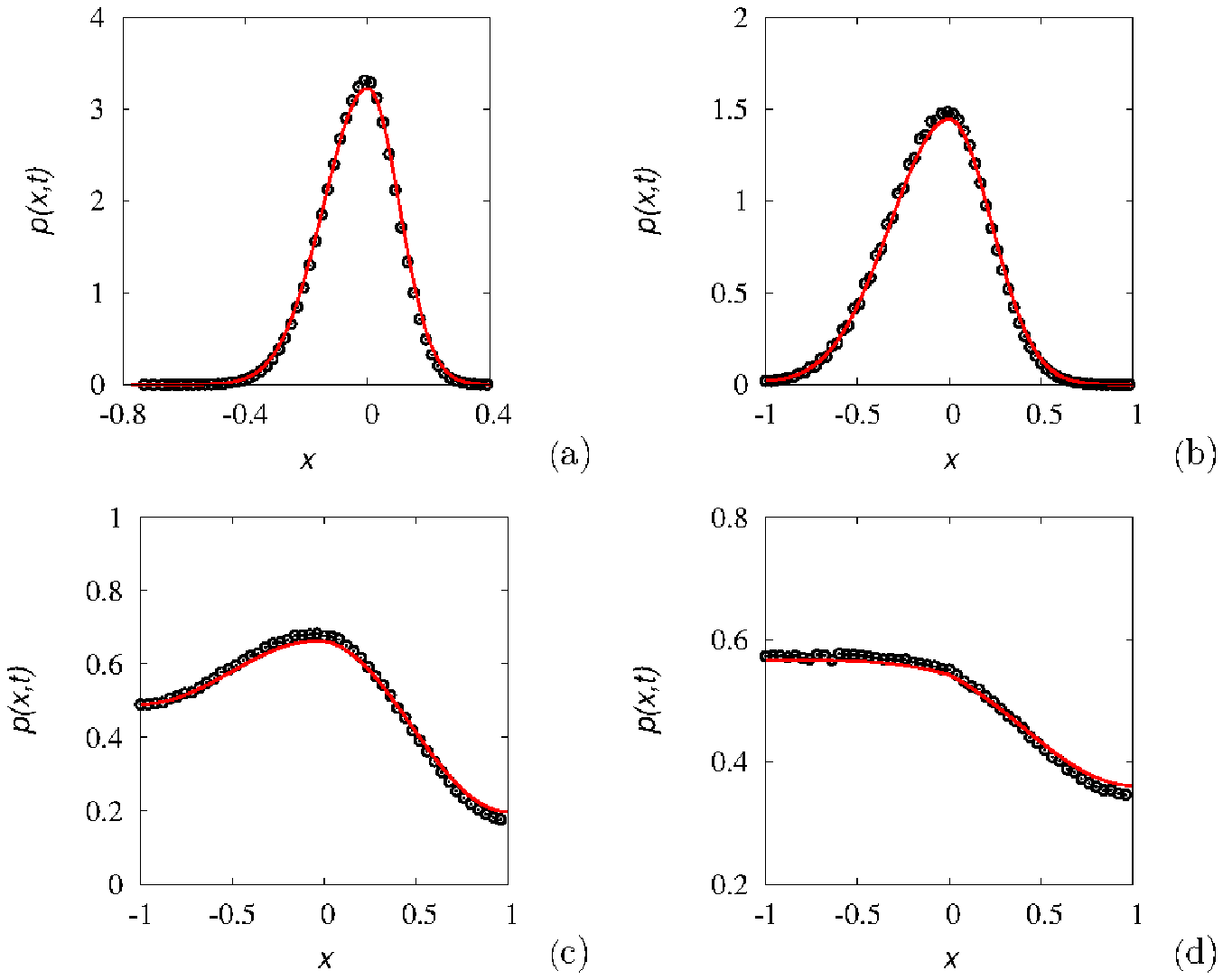}
\end{center}
\caption{
 Same as in figure 2 at $\delta_2=\delta_1/2$, $\tau_2=1/2$.}
\label{Fig3}
\end{figure}
\begin{figure}
\begin{center}
\includegraphics[width=12cm]{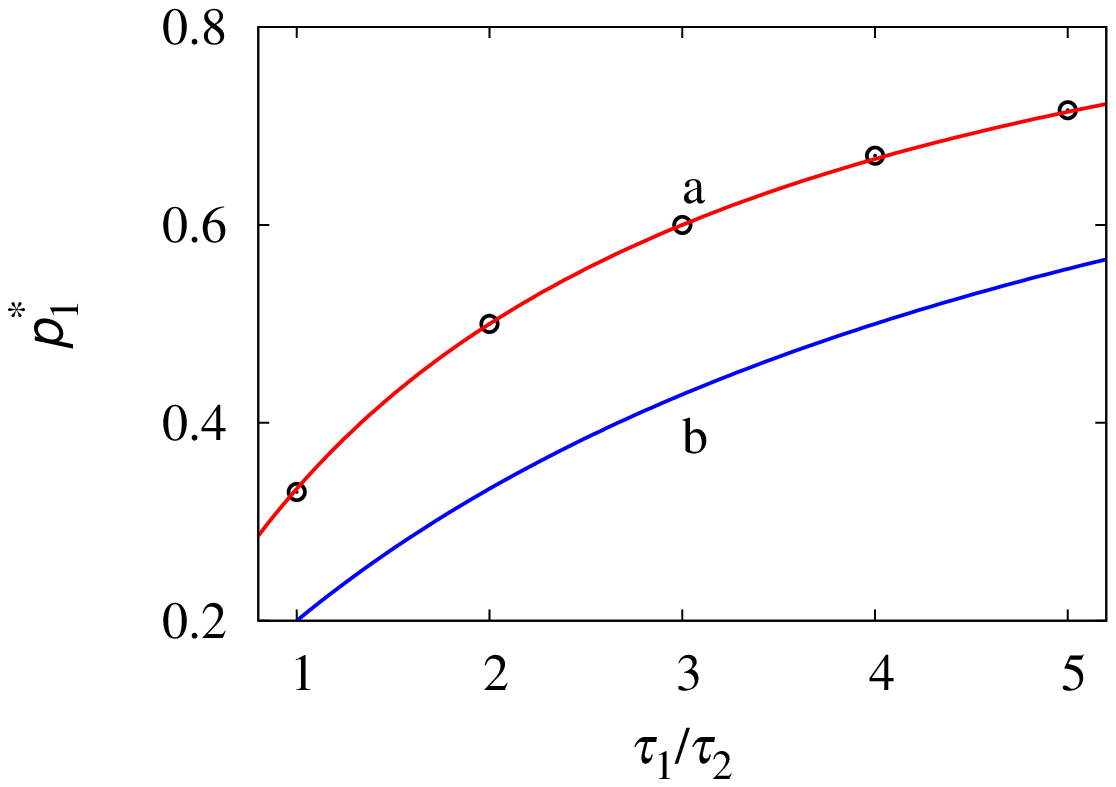}
\end{center}
\caption{Particle frequency $p_1^*$ in phase ``1'' at steady state
as a function of  $\tau_1/\tau_2$ in
a MuPh-LRW closed cell $x \in [-1,1]$ at $\delta_2=\delta_1/2$.
 Symbols are the results of lattice simulations,
curve (a) corresponds to $p_1^*=(b_2/b_1) \, p_2^*$,
curve (b) to $p_1^*=(D_2/D_1) \, p_2^*$.}
\label{Fig4}
\end{figure}

The lattice simulation results are accurately described by the hyperbolic hydrodynamic
model, and the validity of the velocity-based interfacial
condition (\ref{eq4}), or eq. (\ref{eq3}), is highlighted 
 by the data depicted in figure \ref{Fig4} referred to 
the particle fraction $p_1^*$ at steady state in phase   ``1'',
obtained from lattice simulations, as a function of the ratio $\tau_1/\tau_2$, compared with
the result deriving from eq. (\ref{eq4}) and constrasted with the
parabolic interpretation of the boundary conditions
based on the ratio of the phase diffusivities.

\begin{figure}
\begin{center}
\includegraphics[width=9cm]{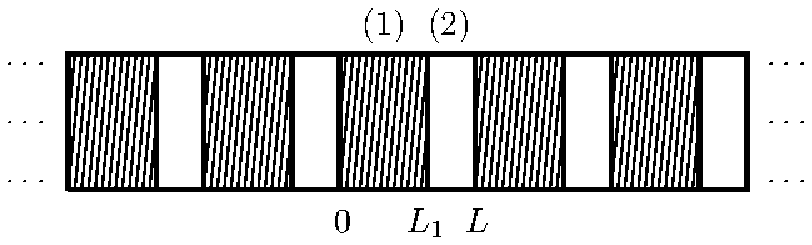}
\end{center}
\caption{Schematic representation of a LRW in a periodic
lattice structure possessing a multiphase unit periodicity
cell.}
\label{Fig5}
\end{figure}

\begin{figure}
\begin{center}
\includegraphics[width=12cm]{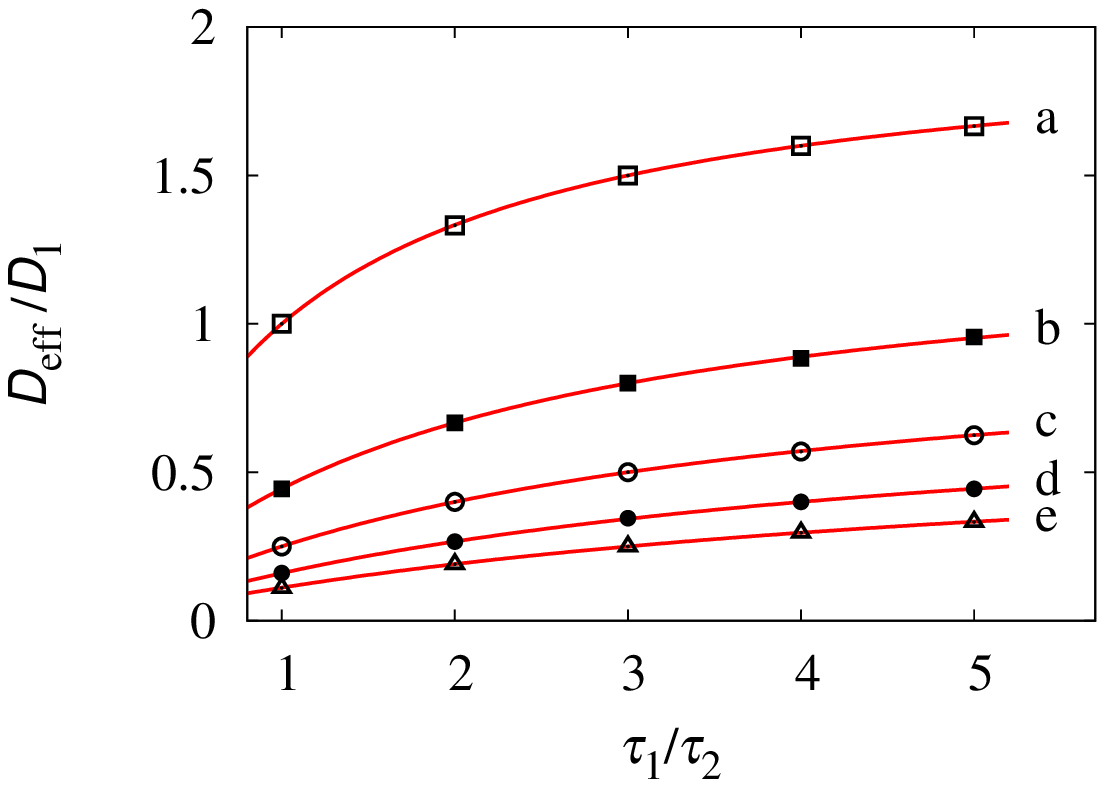}
\end{center}
\caption{Dispersion coefficient $D_{\rm eff}/D_1$
for particle transport in a periodic multiphase lattice
as a function of $\tau_1/\tau_2$.
Symbols are the results of lattice simulations, solid lines the
predictions of homogenization theory applied to the hyperbolic
transport model. Line (a) and ($\square$) refers to $\delta_1/\delta_2=1$,
line (b) and ($\blacksquare$)  to $\delta_1/\delta_2=2$,
line (c) and ($\circ$) to $\delta_1/\delta_2=3$, line (c) and ($\bullet$) 
to $\delta_1/\delta_2=4$, line (e) and ($\triangle$) to 
$\delta_1/\delta_2=5$.}
\label{Fig6}
\end{figure}

Once the qualitative and quantitative validity  of the
hyperbolic description and of the interfacial conditions
arising from it has been assessed, it is possible to use
this hydrodynamic model to investigate finer transport properties
of MuPh-LRW. Specifically, we consider  a dispersion experiment
on a lattice composed by the periodic repetition of a multiphase
unit cell as depicted in figure \ref{Fig5}.
The unit lattice structure is the same used for the data in
figures 2 and \ref{Fig3}, with physical length $L=2$,
$L_1=L_2=1$ and $\delta_1=1/N$, where $N=N_1=100$,
$\tau_1=1$, $\delta_2=\delta_1/\alpha$ and $\tau_2$ varies.
By considering an ensemble of $10^6$ particles initially
located at $z_0=0$ ($x_0=0$), the first order moments (mean and mean square
displacement) are estimated, and from their linear scalings with time $t$ in the
long-term regime, the value of the effective velocity $V_{\rm eff}$
and effective diffusivity $D_{\rm eff}$ (dispersion coefficient) obtained.
From the simulations one obtains $V_{\rm eff}=0$, while the results for 
$D_{\rm eff}$ as a function of $\tau_1/\tau_2$, varying $\delta_1/\delta_2$
are depicted in figure \ref{Fig6}.
 
\begin{figure}
\begin{center}
\includegraphics[width=12cm]{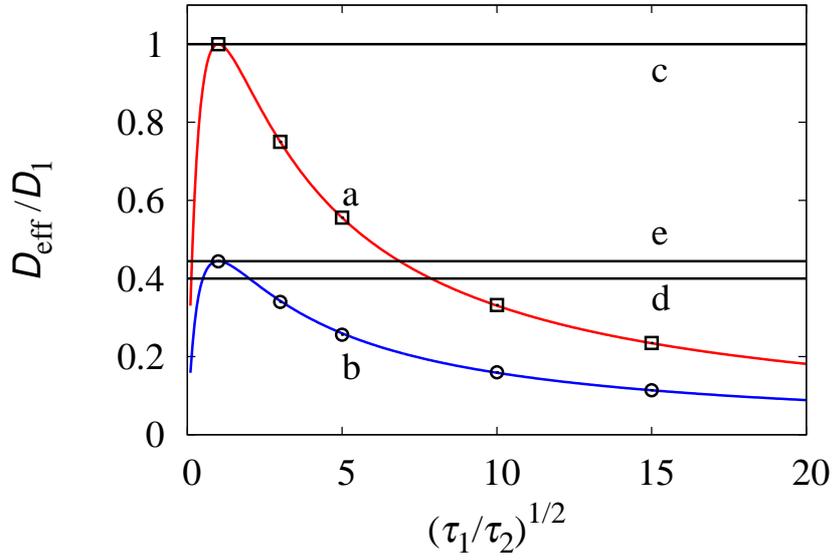}
\end{center}
\caption{$D_{\rm eff}/D_1$ vs $(\tau_1/\tau_2)^{1/2}$
for dispersion in a periodic  multiphase lattice   
for a fixed  ratio
$D_1/D_2$. Symbols are the results
of lattice simulations, solid lines (a) and (b) the
predictions of the hyperbolic transport model eq. (\ref{eq5}).
Line (a) and ($\square$) refer to $D_1/D_2=1$,
line (b) and ($\circ$) to $D_1/D_2=4$.
The horizontal lines represent the lower and upper
bounds in the predictions of the parabolic
transport model obtained by varying $\lambda$: line (c) refers to
$D_1/D_2=1$, line (d) and (e) to $D_1/D_2=4$.}
\label{Fig7}
\end{figure}

These data should be compared with the long-term properties
derived from the continuous hyperbolic model based on eq. (\ref{eq1})
obtained from exact moment analysis \cite{giona_pre}.
In order to have a qualitative picture of the influence of the
finite velocity assumption in the long-term hydrodynamic behavior of
MuPh-LRW, we consider also the modeling of particle
motion in terms of a Langevin equation driven by a Wiener process,
and leading to a parabolic Fokker-Planck equation. Taking
into account the nonlinear and discontinuous nature of the resulting
Langevin equation,  we consider the most general interpretation
of it, namely $d x(t)= \sqrt{2 \, D(x(t))} *_\lambda d w(t)$,
where $D(x)=D(x+L)$ is the discontinuous and spatially
periodic diffusivity profile attaining the values $D_h$ in each lattice
phase, $d w(t)$  the increment of a Wiener process in the time
interval $dt$ and ``$*_\lambda$'' indicates that the $\lambda$-calculus,
$\lambda \in [0,1]$, has been chosen in the definition
of the stochastic Stieltjes integrals ($\lambda=0,\,1/2,\,1$ correspond
to the Ito, Stratonovich and H\"anggi-Klimontovich interpretation, 
respectively) 	\cite{kloeden}. A detailed account
 of the exact homogenization analysis
of the different hydrodynamic models can be found in \cite{giona_pre}.
The final result (for $L_1=L_2=1$) is 
\begin{equation}
\frac{1}{D_{\rm eff}}= \frac{1}{2} \left ( \frac{1}{b_1}+
\frac{1}{b_2} \right ) \left ( \frac{1}{b_1 \, \tau_1}+
\frac{1}{b_2\, \tau_2} \right ) 
\label{eq5}
\end{equation}
for the hyperbolic model, and
\begin{equation}
\frac{1}{D_{\rm eff}}= \frac{1}{4} \left ( \frac{1}{D_1^{1-\lambda}}
+ \ \frac{1}{D_2^{1-\lambda}} \right )
\left ( \frac{1}{D_1^{\lambda}}
+ \ \frac{1}{D_2^{\lambda}} \right )
\label{eq6}
\end{equation}
for the Langevin-Wiener model associated with a $\lambda$-interpretation
of the stochastic integrals. The data depicted in figure \ref{Fig6}
clearly indicate that the hyperbolic transport
model accurately accounts for the dispersion
properties in a multiphase lattice.
Conversely, even keeping $\lambda$ as an adjustable 
parameter, it is impossible for any Langevin-Wiener model of MuPh-LRW diffusion
to provide a quantitative estimate of the
long-term dispersion properties. This claim  is supported
by the data depicted in figure \ref{Fig7}. These
data refer to the effective diffusion coefficient in
a periodic multiphase lattice at $\delta_2=\delta_1/\gamma \,  \xi$,
$\tau_2=\tau_1/\xi^2$ by varying the parameters $\gamma$, and $\xi$.
For fixed values of $\gamma$, the ratio
$D_1/D_2=\gamma$ is constant. From eq. (\ref{eq6}) it follows
that any Langevin-Wiener model of particle transport would
predict a constant value of $D_{\rm eff}$ independently of
the value of $\xi$. 
Conversely, the hyperbolic
 model based on eq. (\ref{eq1}) predicts
a value of $D_{\rm eff}/D_1$ that depends continuously on the
ratio $\tau_1/\tau_2=\xi^2$ for fixed $\gamma$. Lattice
simulation results depicted in figure \ref{Fig7} support
the latter  prediction of the hyperbolic
hydrodynamic model.
In the case $D_1/D_2=1$, one has $D_{\rm eff}/D_1=1$ from
the parabolic model, independently of $\lambda$ (line (c) in figure \ref{Fig7}),
while in general $D_{\rm eff}/D_1$in parabolic Langevin-Wiener models
 is lower- and upper-bounded by the
values attained at $\lambda=0$ and $\lambda=1/2$ (see the Appendix).

 This  result indicates
that the  hyperbolic hydrodynamic model not only provides
a more quantitatively consistent alternative to parabolic
models for describing LRW at
short timescales, as addressed in \cite{giona_lrw}, but 
it is the only continuous model deriving from a continuous
stochastic description of particle transport constistent with
long-term dispersion data in multiphase periodic lattices.

Reversing the  latter argument, it implies that the
assumption of
finite-time propagation velocity is a fundamental prerequisite
in order to predict correctly both the long-time dispersion
properties in infinite multiphase lattices and the equilibrium properties
in closed multiphase cells.

The latter observation opens up interesting perspectives in the
hydrodynamic modeling of particle systems based on the fundamental
assumption of finite propagation velocity. This observation
finds a significant experimental confirmation in the
ubiquitous evidence of ballistic transport at short timescales
both in micro- and nanostructures \cite{ball1,ball2,ball3},
and sheds new light on the conceptual relevance in non-equilibrium
statistical physics of stochastic approaches deeply grounded
on the ``weak relativistic principle'' of finite propagation
velocity. The same approach can be  extended
to electron-transport in periodic lattices, in order to
determine the effective electron mass in all the
solid-state systems in which  experimental evidence
suggests an  effective relativistic constraint associated with
a bounded velocity of carrier particles \cite{geim1,geim2}.

\section*{Appendix}
These notes address some auxiliary  observations and explanations
 complementing the results developed
in the main text.

\subsection*{LRW and Hyperbolic Hydrodynamic Models}
\label{sec1}
In \cite{giona_lrw} an hyperbolic hydrodynamic model for the classical
asymmetric Lattice Random Walk (LRW) has been derived.
Three parameters characterize a LRW:
(i) $\delta$ the distance between
nearest neighboring sites, (ii) $\tau$ the hopping time, and  (iii) $r=r_1-r_2$
the difference between the probabilities of moving to the
right ($r_1$) and  the left ($r_2$),  where $r_1,\, r_2 >0$,
$r_1+r_2=1$.
In the symmetric case, $r=0$, the hyperbolic model for the
partial probability waves $p_\pm(x,t)$ reads
\begin{eqnarray}
\frac{\partial p_+(x,t)}{\partial t} & = & - b \, \frac{\partial p_+(x,t)}{\partial
x} - \lambda \, \left [ p_+(x,t)-p_-(x,t) \right ] \nonumber \\
\frac{\partial p_-(x,t)}{\partial t} & = &  b \, \frac{\partial p_-(x,t)}{\partial
x} + \lambda \, \left [ p_+(x,t)-p_-(x,t) \right ] 
\label{aeq1}
\end{eqnarray}
where $b=\delta/\tau$ is the finite lattice velocity.
Regarding the relation between $\lambda$ and $\tau$ in the symmetric
case, it is important to observe the following.
The statistical description defined by eq. (\ref{aeq1}) can be
interpreted in the symmetric case in two ways, in the
meaning that two different Generalized Poisson-Kac (GPK)  processes
\cite{gpk1} give rise to the same statistical description.

A GPK process
 possessing $N=2$ stochastic states on the real
line ${\mathbb R}$, is defined by  a system
of velocities $b_1=b$, $b_2=-b$, referred to the two stochastic
states, a vector of transition
rates $\boldsymbol{\Lambda}=(\lambda_1,\lambda_2)$ and a
transition probability matrix ${\bf A}=(A_{\alpha,\beta})_{\alpha,\beta=1}^2$
\cite{gpk1}.
The latter two quantities define the 2-state finite Poisson
process $\chi_2(t;{\boldsymbol \Lambda},{\bf A})$ that is essentially
a 2-state Markov chain, so that the dynamics is specified
by the stochastic differential equation
\begin{equation}
d x(t)= b_{\chi_2(t;{\boldsymbol \Lambda},{\bf A})} \, d t
\label{aeq2}
\end{equation}
In the application of GPK models to LRW, the transitions rates
are equal, i.e., $\lambda_1=\lambda_2=\lambda$.

The first GPK model corresponds to the choice $\lambda=2/\tau$ and
\begin{equation}
{\bf A}= \frac{1}{2} \, 
\left (
\begin{array}{ll}
1 & 1 \\
1 & 1 
\end{array}
\right )
\label{aeq3}
\end{equation}
which corresponds to the assumption of equal probabilities
of selecting one of the two stochastic states at any transition
point. Observe that the definition of ``stochastic
state'' adopted here for a GPK process has  nothing to share
with the concept of a lattice phase introduced in the main
article for MuPh-LRW.
The above GPK model corresponds to the approach followed in
\cite{giona_lrw}.
The second alternative is to choose
\begin{equation}
\lambda=\frac{1}{\tau}
\label{aeq4}
\end{equation}
defining the transition probability matrix as
\begin{equation}
{\bf A}=  
\left (
\begin{array}{ll}
0 & 1 \\
1 & 0 
\end{array}
\right )
\label{aeq5}
\end{equation}
in which at any transition point there is a switching to
the other stochastic state (implying a reversal of velocity direction).
This is the approach followed in the main text. In the latter case,
eq. (\ref{aeq2}) reduces to the classical Poisson-Kac model
\begin{equation}
d x(t)= b \, (-1)^{\chi(t;\lambda)} \, d t
\label{aeq6}
\end{equation}
where $\chi(t;\lambda)$ is a classical Poisson process possessing
transition rate $\lambda$.

\subsection*{Computational issues}
\label{sec2}

In the simulations of a lattice multiphase unit cell reported in
the main text we have chosen $\delta_1=1/N$ with $N=100$,
$\tau_1=1$, letting $\delta_2$ and $\tau_2$ vary.
Specifically, $\delta_2=\delta_1/\alpha$ where $\alpha=1,2,\dots$ is an
integer. Since the multiphase cell has been
 defined in the interval $x \in [-1,1]$, this corresponds to consider
$N_1=N$ sites for the lattice phase ``1'', defined
for $x \in [-1,0)$ and $N_2=\alpha \, N$ sites
for phase ``2'' defined for $x \in (0,1]$.

\begin{figure}
\begin{center}
\includegraphics[width=16cm]{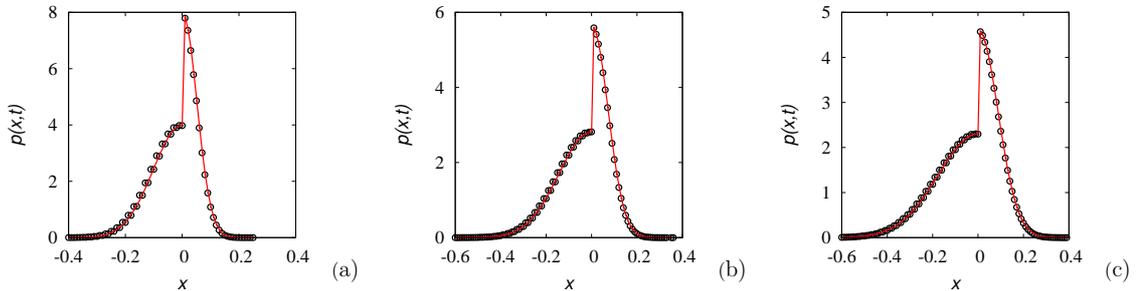}
\end{center}
\caption{$p(x,t)$ vs $x$ for a MuPh-LRW in a closed cell $x \in [-1,1]$
at $\delta_2/\delta_1=1/2$, $\tau_2=\tau_1$ starting from an impulsive initial condition centered at the interfacial point. Symbols ($\bullet$)
refer to simulations with $N=100$, $\tau_1=1$, lines to
$N=300$, $\tau_1=1/9$. Panel (a) to (c) correspond to
$t=100,\,200,\,300$, respectively.}
\label{AFig1}
\end{figure}

The results obtained from stochastic  lattice simulations of MuPh-LRW
are not influenced by a finer representation of the lattice
obtained by increasing $N$, if the timescale $\tau_1$ is properly
rescaled. The most critical issue is related to the short-time
behavior. Figure \ref{AFig1} depicts the comparison of two
lattice experiments on a closed MuPh-LRW cell for
two values of $N=100,\,300$. In the latter case, the value
of $\tau_1$ as been rescaled to $\tau_1=1/9$.

There are no significant differences in the two simulations other than
that the case $N=300$ obviously provides density profiles there are
slightly more smooth.

A further comment refers to the initial conditions
in a closed multiphase cell. We assumed that the initial
condition is centered at the interface. In order to
avoid unpleasent parity-effects associated with lattice random walk
(if the initial condition is at an even site, then the
concentration evaluated at even lattice times is vanishing at all the
odd sites, and viceversa), the initial conditions has been distributed
equally at site $z_0=0$ and at the first neighboring site belonging
to phase ``2'', i.e., $z_1=1$. In the solution of the
continuous hyperbolic model this has been taken into account by considering
an initial condition uniformly distributed within the interval $x \in [0,\delta_2]$.
Moreover, in the  simulations of the
hyperbolic continuous model it has been assumed
an initial  balancing of the two partial probability waves,
namely
$p_+(x,0)=p_-(x,0)$.

\begin{figure}
\begin{center}
\includegraphics[width=16cm]{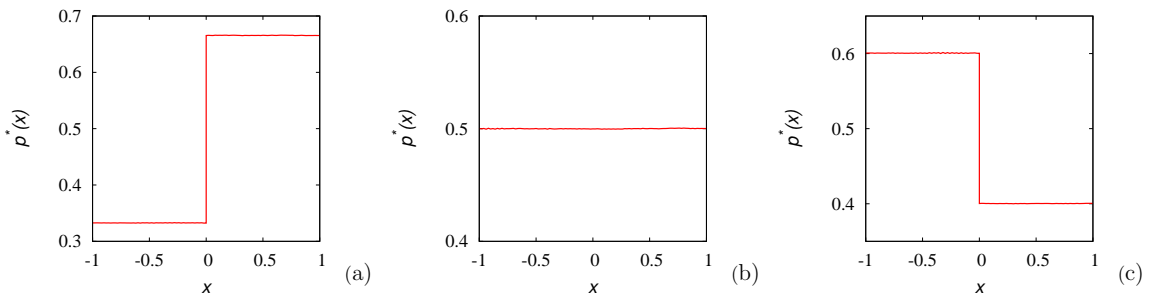}
\end{center}
\caption{Stationary concentration profiles $p^*(x)$ vs $x$
at $\delta_1/\delta_2=2$. Panel (a) refers to $\tau_2=\tau_1$,
(b) to $\tau_2=\tau_1/2$, (c) to $\tau_2=\tau_1/3$.}
\label{AFig2}
\end{figure}

The lattice simulation data reported in the main text refer to the
long-term behavior of a closed MuPh-LRW system, once the dynamics
has relaxed towards the equilibrium distribution $p^*(x)$. Figure \ref{AFig2}
depicts the simulations results obtained for $p^*(x)$.
Since the number of lattice particles considered is $N_p=10^6$, in
order to  obtain a  statistically  accurate
estimate  of the equilibrium
distributions, reducing the fluctuations associated with
finite-size effects  in the particle ensemble,
the stationary concentration profiles
have been averaged over $N_t=10^4$ time instants, after  reaching
 equilibrium conditions.
The results of this averaging  is depicted in figure \ref{AFig2}
for $\delta_2=\delta_1/2$ at three  different values of $\tau_2/\tau_1$.

\begin{figure}
\begin{center}
\includegraphics[width=10.5cm]{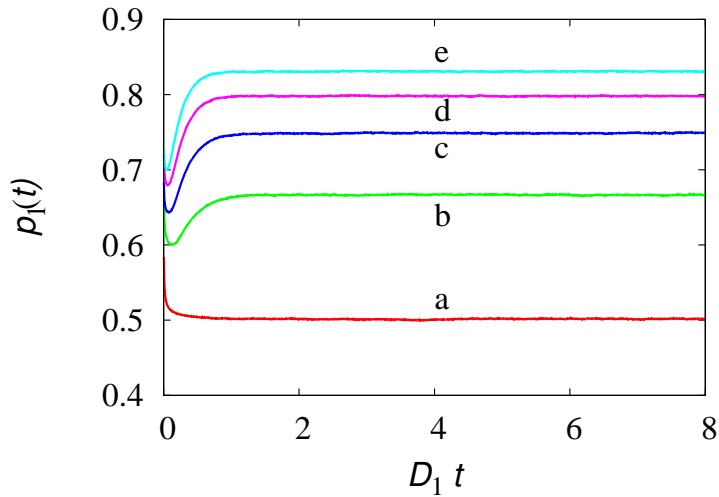}
\end{center}
\caption{$p_1(t)$ vs $t$ in a closed MuPh-LRW cell at $\delta_2=\delta_1$
for several values of $\tau_2$: from (a) to (e), $\tau_2=\tau_1/h$, $h=1,2,3,4,5$.}
\label{Xfigure3}
\end{figure}

Indicate with $p_1(t)$ the fraction of particles
at time $t$ within phase ``1''. The dynamics of this
quantity, relaxing towards $p_1^*$ (depicted in figure  \ref{Fig4} of 
the main text),
is shown in figure \ref{Xfigure3} for $\delta_2=\delta_1$, at several
values of $\tau_2$.
The data have been reported using the normalized abscissa $D_1 \, t$.
Since $\delta_1=1/N$, $N=100$,  $\tau_1=1$, $D_1=5 \times 10^{-5}$ a.u.

\subsection*{Boundary conditions}
\label{sec3}

The boundary conditions for MuPh-LRW in the presence of ideal interfaces
can be derived in several different ways.
In a fully lattice description, MuPh-LRW is solely a simple
symmetric walk on ${\mathbb Z}$ parametrized with respect to the
lattice time $n \in {\mathbb N}$. Its statistical description
involves the probabilities $P_h^n$ of finding the particle at the
lattice site $h$ at the lattice time $n$, fulfilling the
Markov dynamics $P_n^{h+1}= (P_{h-1}^n+P_{h+1}^n)/2$.

In a spatially continuous representation of the process, when
positions $ x \in {\mathbb R}$, the spatial heterogeneity,
associated with the different values of $\delta_h$ in the
two lattice phases plays a role. Let $p(x,n)$ the probability density function
continuously parametrized with respect to the spatial coordinate $x$.
The probability $P_h^n$ corresponds to the integral of
the continuous $p(x,n)$ over an interval centered at $x_h$ of
width $\delta_{\sigma(x_h)}$, where $\sigma(x_h)=1,2$,
depending whether the site belongs to phase ``1'' or ``2''.,
Consequently,
\begin{equation}
P_h^n \simeq p(x_h,n) \, \delta_{\sigma(x_h)}
\label{eq3a_1}
\end{equation}
The hyperbolic stochastic model associated
with $p(x,n)$, taking a continuation of $n$ towards real values,
stems from a local stochastic dynamics given by

\begin{equation}
d x(n)= \delta(x) \, (-1)^{\chi(n,1)} \, d n
\label{eq3a_2}
\end{equation}
where $\delta(x)=\delta_1$ if $\sigma(x) =1$, and
$\delta(x)=\delta_2$ if $\sigma(x)=2$, and the transition rate
equals $1$.

Henceforth, let $\Omega_1$, $\Omega_2$ be the subset of
${\mathbb R}$, $\Omega_1 \cup \Omega_2 = {\mathbb R}$ , occupied
by phase ``1'' and phase ``2'', respectively, and
let $p_\pm^{(h)}(x,n)$ the probability densities
in the two lattice phases, $h=1,2$.

The hyperbolic  hydrodynamic model expressed with respect to the
continuation of the lattice time $n$ towards real values  is given
by
\begin{equation}
\frac{\partial p_\pm^{(h)}(x,n)}{\partial n} =  \mp \delta_h \, \frac{\partial
p_\pm^{(h)}(x,n)}{\partial x} \mp \left [p_+^{(h)}(x,n)-p_-^{(h)}(x,n) \right ]
\qquad x \in \Omega_h
\label{eq3a_3}
\end{equation}
Next, account for the time subordination of $t$ with respect to $n$,
\begin{equation}
d t = \tau(x) \, d n
\label{eq3a_4}
\end{equation}
where
\begin{equation}
\tau(x)=
\left \{
\begin{array}{ccc}
\tau_1 & & \in \Omega_1 \\
\tau_2 & & \in \Omega_2 
\end{array}
\right .
\label{eq3a_5}
\end{equation}
which, in the present case of hyperbolic stochastic dynamics, means
\begin{equation}
\frac{\partial }{\partial n}= \frac{d t}{d n} \,
 \frac{\partial }{\partial t}
\label{eq3a_6}
\end{equation}
Consequently,  with respect to the physical time $t$, the
balance equations (\ref{eq3a_3}) becomes
\begin{eqnarray}
\frac{\partial p_\pm^{(1)}(x,t)}{\partial t} & = &
 \mp b_1 \frac{\partial p_\pm^{(1)}(x,t)}{\partial t} \mp \lambda_1
\, \left [p_+^{(1)}(x,t)-p_-^{(1)}(x,t) \right ] \qquad x \in \Omega_1
\nonumber \\
\frac{\partial p_\pm^{(2)}(x,t)}{\partial t} & = &
 \mp b_2 \frac{\partial p_\pm^{(2)}(x,t)}{\partial t} \mp \lambda_2
\, \left [p_+^{(2)}(x,t)-p_-^{(2)}(x,t) \right ] \qquad x \in \Omega_2
\label{eq3a_7}
\end{eqnarray}
where $b_h=\delta_h/\tau_h$, $\lambda_h=1/\tau_h$,
which is the model considered in the main text.
The boundary condition at an ideal interface follows from
the hyperbolic structure of eq. (\ref{eq3a_7}) by
imposing that there is no singularity at the interface
(alternatively, one can invoke steady-state conditions).
To this purpose,  instead of eq. (\ref{eq3a_7}) let us
consider a mollified version of it, by
defining a smooth velocity field $b(x;\varepsilon)$,
which is $C^k({\mathbb R})$, with $k \geq 1$ with respect to $x$
for any $\varepsilon>0$, and such that, in the
limit for $\varepsilon \rightarrow 0$, it reproduces the
discontinuity in the lattice phase velocities,
\begin{equation}
\lim_{\varepsilon \rightarrow 0} b(x;\varepsilon)
= \left \{
\begin{array}{ccc}
b_1 & & x \in \Omega_1 \\
b_2 & & x \in \Omega_2
\end{array}
\right .
\label{eq3a_8}
\end{equation}
and similarly for the transition rates $\lambda_h$, introducing a smooth
field  $\lambda(x;\varepsilon)$.
With respect to this mollified description,
eq. (\ref{eq3a_7}) becomes
\begin{equation}
\frac{\partial p_\pm(x,t)}{\partial t}
= \mp \frac{\partial \left [ b(x;\varepsilon) \, p_\pm(x,t) \right ]}{\partial x}
- \lambda(x;\varepsilon) \, \left [p_+(x,t)-p_-(x,t) \right ]
\label{eq3a_9}
\end{equation}
Let $x_0$ be the position of an ideal interface.
Integrating eq. (\ref{eq3a_9}) in the interval
$[x_0-\eta,x_0+\eta]$, where $\varepsilon >0$ is a small parameter,
one obtains
\begin{eqnarray}
\pm  \left [ b(x_0+\eta;\varepsilon) \, p_\pm(x_0+\eta,t) -
b(x_0-\eta;\varepsilon) \, p_\pm(x_0-\eta,t) \right ]
=
- \int_{x_0-\eta}^{x_0+\eta} \frac{\partial p_\pm(x,t)}{\partial t} \, d x
\nonumber \\
- \int_{x_0-\eta}^{x_0+\eta}  \lambda(x;\varepsilon) \, \left [p_+(x,t)-p_-(x,t) \right ] \, 
\label{eq3a_10}
\end{eqnarray}
and let
\begin{eqnarray}
Dp_{\rm max} &  = & \max_{x \in [x_0-\eta,x_0+\eta]}
\left | \frac{\partial p_\pm(x,t)}{\partial t} \right |
\, , \qquad
\lambda_{\rm max}= \max_{x \in [x_0-\eta,x_0+\eta]}
  \lambda(x;\varepsilon) \
\nonumber \\
\Delta_{\rm max} & = & \max_{x \in [x_0-\eta,x_0+\eta]}
\left |p_+(x,t)-p_-(x,t) \right |
\label{eq3a_11}
\end{eqnarray}
All these quantities can be assumed to be finite, at least for sufficiently
long timescales. Consequently,
\begin{equation}
|b(x_0+\eta;\varepsilon) \, p_\pm(x_0+\eta,t) -
b(x_0-\eta;\varepsilon) \, p_\pm(x_0-\eta,t)| 
\leq 2 \, \eta \left [ Dp_{\rm max} + \lambda_{\rm max} + \Delta_{\rm max}
\right ] 
\label{eq3a_12}
\end{equation}
and, in the limit for  $\eta \rightarrow 0$, one
recovers
\begin{equation}
b(x_0^-; \varepsilon) \, p_\pm(x_0^-,t) =
b(x_0^+; \varepsilon) \, p_\pm(x_0^+,t)
\label{eq3a_13}
\end{equation}

Taking the limit for $\varepsilon\rightarrow 0$,
the velocity-based boundary condition follows
\begin{equation}
b_1 \, p_\pm^{(1)}(x_0,t) = b_2 \, p_\pm^{(2)}(x_0,t)
\label{eq3a_14}
\end{equation}
introduced in the main text. 

It is possible to provide a lattice-dynamics  based analysis  of
the interfacial boundary conditions leading to
eq. (\ref{eq3a_14}) generalizing the approach developed by
Ovaskainen and Cornell \cite{ovas} which derive the same boundary
condition in Appendix A.1 Case A (in the case of pure spatial
heterogeneity). In point of fact, in the Ovaskainen and Cornell
paper, the symmetric case of an ideal interface
considered in the present work corresponds
to $z=0$ in their calculations, and the quantities
$q_\pm$ used in their analysis are proportional to $\delta_h$, $h=1,2$.

\subsection*{Dispersion in a periodic MuPh-LRW structure}
\label{sec4}

The data for the effective diffusion  coefficient $D_{\rm eff}$ reported
in the main text, refer to the long-term emergent behavior
observed in particle motion in a periodic multiphase
lattice. The unit cell considered is the
same used for closed-system numerical experiments, consisting of
$N_1=100$ sites for phase ``1'' and $N_2= \alpha \, N_1$ sites
for phase ``2'' where $\alpha$ is defined by $\delta_2=\delta_1/\alpha$.

\begin{figure}
\begin{center}
\includegraphics[width=12cm]{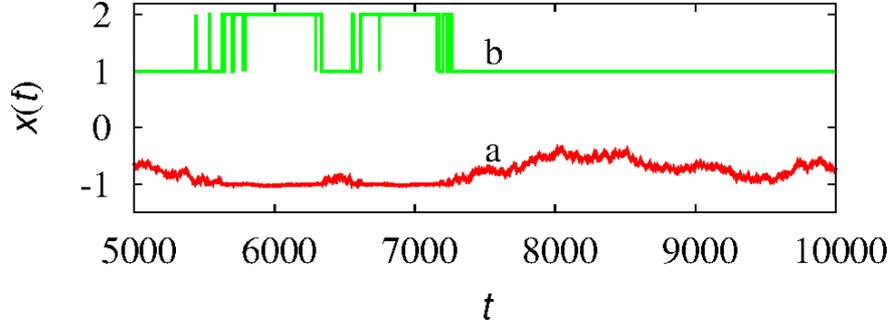}
\end{center}
\caption{Portion of a trajectory (line a) of a lattice particle in a multiphase
periodic lattice at $\delta_2=\delta_1/10$, $\tau_2=\tau_1$.
Line (b) indicates the phase of the site at which the particle
is located at time $t$.}
\label{AFig0}
\end{figure}

Figure \ref{AFig0} depicts a portion of a trajectory of a lattice
particle moving in a multiphase periodic lattice,
in the rather extreme situation $\delta_2=\delta_1/10$,
$\tau_2=\tau_1$.

\begin{figure}
\begin{center}
\includegraphics[width=10.5cm]{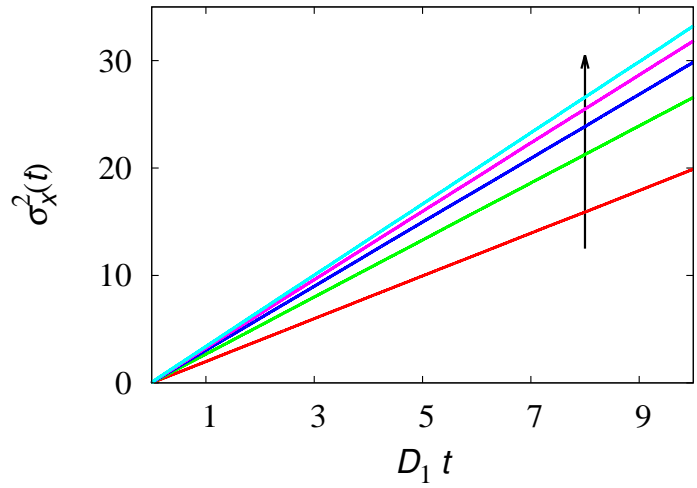}
\end{center}
\caption{Mean square displacement $\sigma_x^2(t)$ vs $D_1 t$ for
 particle motion in a MuPh-LRW periodic lattice at $\delta_2=\delta_1$
for several values of $\tau_2=1/h$, $h=1,\dots,5$, ($\tau_1=1$). The arrow
indicates increasing values of $h$.}
\label{AFig3}
\end{figure}

The results for $D_{\rm eff}$ stem from the long-term
linear scaling with time
of the mean square displacement $\sigma_x^2(t)$,  that fulfils
the emergent Einsteinian behavior
\begin{equation}
\sigma_x^2(t) \sim  2 \, D_{\rm eff} \, t \qquad D_1 t \gg 1
\label{eqz_1}
\end{equation}
as depicted in figure \ref{AFig3} at $\delta_2=\delta_1$ for
several values of the ratio $\tau_2/\tau_1$.

The continuous hydrodynamic model for MuPh-LRW in periodic lattices
stems from a space-time continuous description of
stochastic particle motion in the presence of ideal interfaces.
Since an ideal interface does not exert any specific active
contribution to random particle dynamics, but neutrally separates
the two lattice phases, it is intuitive that continuous hydrodynamic models
should be derived from simple stochastic motion
on ${\mathbb R}$ in a time-continuous setting, in which the
heterogeneity in transport parameters, deriving from the multiphase
partition of the lattice, is accounted for.

As discussed in the previous paragraphs, the continuous
hyperbolic hydrodynamic
model is just the evolution equation for the statistical
description,  expressed by mean of the partial probability waves,
of   the Poisson-Kac process
\begin{equation}
d x(t)=b(x(t)) \, (-1)^{\chi(t;\lambda(x(t)))} \, dt
\label{eqz_2}
\end{equation}
where the functions $b(x)$ and $\lambda(x)$ are
expressed by
\begin{equation}
b(x)=
\left \{
\begin{array}{ccc}
b_1=\delta_1/\tau_1 & & x \in \Omega_1 \\
b_2=\delta_2/\tau_2 & & x \in \Omega_2
\end{array}
\right . \,,
\qquad
\lambda(x)=
\left \{
\begin{array}{ccc}
1/\tau_1 & & x \in \Omega_1 \\
1/\tau_2 & & x \in \Omega_2
\end{array}
\right .
\label{eqz_3}
\end{equation}
This model gives rise to the hydrodynamic equations
(\ref{eq3a_7}) equipped with the interfacial
conditions (\ref{eq3a_14}), out of which the expression for the effective
diffusion coefficient reported in the main text,
and here rewritten, follows (assuming equal fractions
of the two lattice phases, i.e., $L_1=L_2$) \cite{giona_pre}
\begin{equation}
\frac{1}{D_{\rm eff}}= \frac{1}{2} \left ( \frac{1}{b_1}+
\frac{1}{b_2} \right ) \left ( \frac{1}{b_1 \, \tau_1}+
\frac{1}{b_2\, \tau_2} \right ) 
\label{eqz_3bis}
\end{equation}

A parabolic model for MuPh-LRW stems from the
statistical properties of a Langevin equation
for particle dynamics
\begin{equation}
d x(t)= \sqrt{2 \, D(x)} *_\lambda d w(t)
\label{eqz_4}
\end{equation}
driven by the increments $dw(t)$ of a Wiener process
in the time interval $dt$, and ``$*_\lambda$'', $\lambda \in [0,1]$,
indicates that a $\lambda$-calculus has been chosen for
 the stochastic integrals (where $\lambda=0$
provides the Ito, $\lambda=1/2$ the Stratonovich, and $\lambda=1$
the H\"anggi-Klimontovich representations).

In eq. (\ref{eqz_4}), $D(x)$ is the local
diffusivity that depends on position due to
phase heterogeneity and periodicity
\begin{equation}
D(x)=
\left \{
\begin{array}{ccc}
D_1=\delta_1^2/2 \,\tau_1 & & x \in \Omega_1 \\
D_2=\delta_2^2/2 \,\tau_2 & & x \in \Omega_2
\end{array}
\right .
\label{eqz_5}
\end{equation}
Consequently eq. (\ref{eqz_4}) is a  highly nonlinear
Langevin equation, with discontinuous coefficients, and
its properties can be studied by considering a mollified
version of it,
\begin{equation}
d x(t)= \sqrt{2 \, \widetilde{D}(x;\varepsilon)} *_\lambda d w(t)
\label{eqz_6}
\end{equation}
by  introducing $\widetilde{D}(x;\varepsilon)$, $\varepsilon>0$
that represents a family of smooth $C^k({\mathbb R})$ fields, $k \geq 1$,
that in the limit for $\varepsilon \rightarrow 0$
converge to the discontinuous diffusivity profile
expressed by eq. (\ref{eqz_5}),
\begin{equation}
\lim_{\varepsilon \rightarrow 0} \widetilde{D}(x;\varepsilon)=
D(x)
\label{eqz_7}
\end{equation}
We are considering all the possible stochastic integrals,
i.e., all the values of $\lambda \in [0,1]$, in order
to test the capability of a Langevin-Wiener stochastic
models to describe the long-term properties
of a MuPh-LRW. Of course, the choice of $\lambda$
implicitly determines the boundary condition at the
interfacial points separating the two lattice phases.

The long term properties of the stochastic motion defined by (\ref{eqz_4})
can be obtained from the homogenization analysis of
the associated Fokker-Planck equation, and provide
the following expression for the
effective diffusivity \cite{giona_pre}
\begin{equation}
\frac{1}{D_{\rm eff}}= \frac{1}{4} \left ( \frac{1}{D_1^{1-\lambda}}
+ \ \frac{1}{D_2^{1-\lambda}} \right )
\left ( \frac{1}{D_1^{\lambda}}
+ \ \frac{1}{D_2^{\lambda}} \right )
\label{eqz_8}
\end{equation}

It has been shown in the main text that
the hyperbolic model provides the correct behavior
of the effective diffusivity  for a MuPh-LRW in the
presence of ideal interfaces over all the range
of lattice parameters, while this is not the
case for the parabolic models based on the
continuous space-time approximation of stochastic
motion via eq. (\ref{eqz_4}).

\begin{figure}
\begin{center}
\includegraphics[width=8cm]{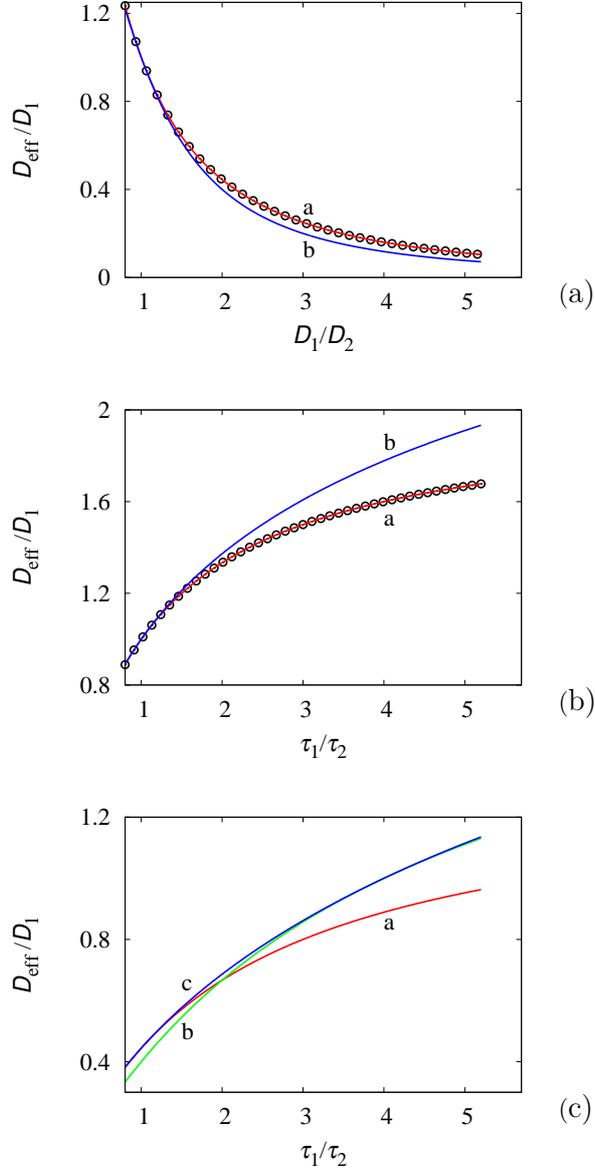}
\end{center}
\caption{Comparison of $D_{\rm eff}/D_1$ for the continuous
models of particle motion in multiphase lattices.
In all the panels, lines (a) correspond to the
results deriving from the hyperbolic model.
 Panel (a) refers to $\tau_2=\tau_1$,
$\delta_2 \neq \delta_1$:
 symbols ($\circ$)  correspond to the
Langevin-Stratonovich, and  line (b) to the Langevin-Ito models,
respectively.
Panel (b) refers to $\delta_1=\delta_2$, in the case
$\tau_2 \neq \tau_1$: symbols ($\circ$) correspond to
the Langevin-Ito, and line (b) to the Langevin-Stratonovich
models, respectively.
Panel (c):  general case, $\delta_2 \neq \delta_1$, $\tau_2 \neq \tau_1$,
in the case
 the ratio of the diffusivity is fixed to
$D_1/D_2=2$. Line (b) corresponds to the
Langevin-Ito, and line (c) to the Langevin-Stratonovich
models, respectively.}
\label{AFig4}
\end{figure}

Figure \ref{AFig4} complements the data shown in the main text.
It provides the comparison of the effective
diffusivity obtained for the hyperbolic model eq. (\ref{eqz_3bis})
(coinciding with that of found in lattice simulations of
MuPh-LRW dispersion), with the diffusivities deriving
from the Langevin-Wiener model eq. (\ref{eqz_8}) at
different values of $\lambda$, for typical conditions.
First of all, observe  from eq. (\ref{eqz_8})
that $D_{\rm eff}|_\lambda= D_{\rm eff}|_{1-\lambda}$,
thus the Ito and the H\"anggi-Klimontovich approaches to
stochastic motion provide the same long-term dispersive behavior.

From the functional structure of the expressions (\ref{eqz_3bis})
and (\ref{eqz_8}) it follows that:
\begin{itemize}
\item if the lattice heterogeneity involves solely the
spatial parameters, i.e., $\tau_1 = \tau_2$, then the
Langevin-Stratonovich approximation provides the
correct expression for the effective diffusion coefficient,
as depicted in panel (a) of figure \ref{AFig4};
\item if the lattice heterogeneity involves the timescales,
i.e., $\delta_1=\delta_2$, the Langevin-Ito (or equivalently
the H\"anggi-Klimontovich interpretation) provides
the right value of  $D_{\rm eff}$ as depicted in panel (b) of figure \ref{AFig4};
\item in the case both spatial and time scales in the two
lattice phases are different, there is no
parabolic model stemming from a Langevin-Wiener approximation
of the lattice dynamics that yields the correct long-term
behavior  found in numerical experiments
of MuPh-LRW. This phenomenon is depicted in panel (c)
of figure \ref{AFig4}.
\end{itemize}

\subsection*{Qualitative analysis}
\label{sec5}
\begin{figure}
\begin{center}
\includegraphics[width=10.5cm]{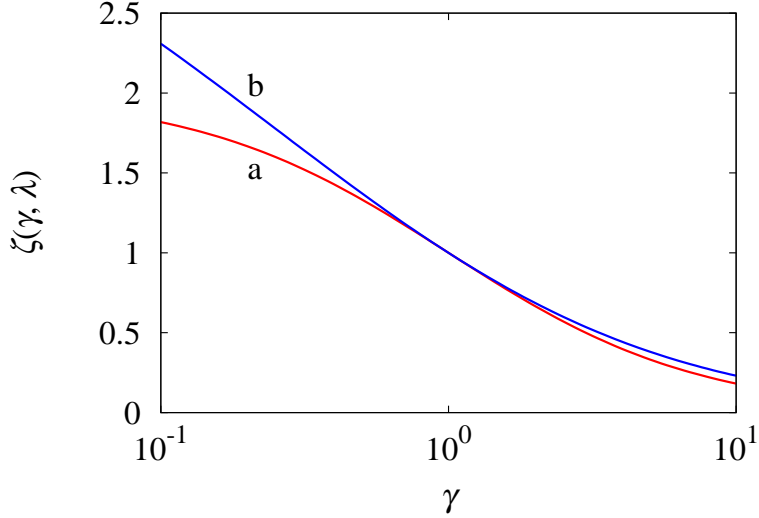}
\end{center}
\caption{Behavior of the function $\zeta(\gamma,\lambda)$
as a function of the ratio $\gamma$ of the two lattice
phase diffusivities. Line (a) refers to $\lambda=0$ (Ito),
line (b) to $\lambda=1/2$ (Stratonovich).}
\label{AFig5}
\end{figure}

\begin{figure}
\begin{center}
\includegraphics[width=10.5cm]{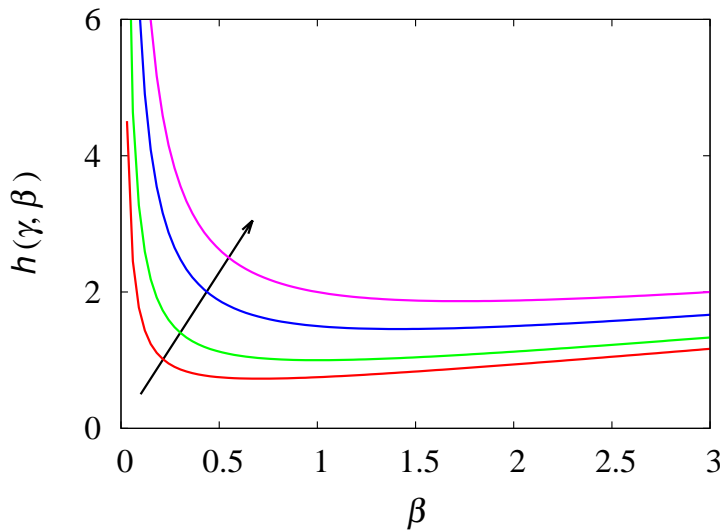}
\end{center}
\caption{$h(\gamma,\beta)$ as a function of $\beta$ for several
values of $\gamma=0.5,\,1,\,2,\,3$. The arrow indicates increasing
values of $\gamma$.}
\label{AFig6}
\end{figure}

It can be useful to explore further the properties
of the expressions for the effective diffusivities.
To begin with, consider the parabolic approximation, i.e.,
eq. (\ref{eqz_8}). In this
case, the ratio $D_{\rm eff}/D_1$ depends solely on
the ratio $\gamma=D_1/D_2$ of the diffusivities
in the two lattice phases
\begin{equation}
\frac{D_1}{D_{\rm eff}} = \frac{1}{4} \left [1+ \gamma^{1-\lambda}
+ \gamma^\lambda + \gamma \right ]= \frac{1}{\zeta(\gamma,\lambda)}
\label{eq5_1}
\end{equation}
The function $\zeta(\gamma,\lambda)$, for fixed $\gamma$, attains
its maximum value at $\lambda=1/2$, and consequently
the Stratonovich interpretation provides
the maximum value of the effective diffusivity, see figure \ref{AFig5}.
Specifically,
\begin{equation}
\zeta(\gamma,0) = \frac{2}{1+\gamma} \, ,
\qquad \zeta \left(\gamma,\frac{1}{2} \right )= \frac{4}{(1+\sqrt{\gamma})^2}
\label{eq5_2}
\end{equation}
Next, consider  the hyperbolic model. Introducing
the nondimensional group $\beta=b_1/b_2$, corresponding to the
ratio of the lattice velocities in the two phases,
eq. (\ref{eqz_3bis}) can be
expressed as
\begin{equation}
\frac{D_1}{D_{\rm eff}}= \frac{1}{4} \left [ (1+\beta)
+ \gamma \left (1+\frac{1}{\beta}  \right ) \right ] = h(\gamma,\beta)
\label{eq5_3}
\end{equation}
The function $h(\gamma,\beta)$ is depicted in figure \ref{AFig6}
as a function of its second argument $\beta$.
The function $h(\gamma,\beta)$, for fixed $\gamma$, attains its
local extremal value at $\beta^*$ (local minimum) given by
\begin{equation}
\beta^*= \sqrt{\gamma}
\label{eq5_4}
\end{equation}
and the corresponding value of $D_1/D_{\rm eff}^*$ is
given by
\begin{equation}
\frac{D_1}{D_{\rm eff}^*}= \frac{(1+\sqrt{\gamma})^2}{4}
\label{eq5_5}
\end{equation}
coinciding with the results obtained
in the Langevin-Stratonovich case.
It follows from this observation that, for fixed $\gamma$,
i.e., for fixed values of the lattice phase diffusivities,
the maximum value of the effective diffusivity in
MuPh-LRW dispersion obtained by varying the ratio $\beta$ between the lattice
velocities is
provided by the Langevin-Stratonovich dispersion coefficient, i.e.,
\begin{equation}
\left . \frac{D_{\rm eff}^{\mbox{MuPh-LRW}}(\beta)}{D_1} \right |_{\mbox{fixed}
\; \gamma}
\leq  \, \frac{D_{\rm eff}^{\mbox{Strato}}(\gamma)}{D_1}
\label{eq5_6}
\end{equation}

\end{document}